\documentclass[preprint,nofootinbib,amsmath,prd,aps,superscriptaddress,tightenlines,12pt]{revtex4}
\usepackage{graphicx}
\usepackage{graphics}
\usepackage{amsmath}
\usepackage{hyperref}

\def\lsim{\mathrel{\rlap{\lower4pt\hbox{\hskip1pt$\sim$}}
    \raise1pt\hbox{$<$}}}                
\def\gsim{\mathrel{\rlap{\lower4pt\hbox{\hskip1pt$\sim$}}
    \raise1pt\hbox{$>$}}}                

\def\OMIT#1{}

\newcommand{\be}{\begin{eqnarray}}
\newcommand{\ee}{\end{eqnarray}}

\newcommand{\nn}{\nonumber}
\newcommand{\w}{\omega}

\newcommand{\bn}{{\bar n}}
\newcommand{\bea}{\begin{eqnarray}}
\newcommand{\eea}{\end{eqnarray}}

\newcommand{\bnp}{\bar n \!\cdot\! p}

\def\lsim{\mathrel{\rlap{\lower4pt\hbox{\hskip1pt$\sim$}}
    \raise1pt\hbox{$<$}}}                
\def\gsim{\mathrel{\rlap{\lower4pt\hbox{\hskip1pt$\sim$}}
    \raise1pt\hbox{$>$}}}                

\def\OMIT#1{}

\begin{document}

\setlength\baselineskip{17pt}

\begin{flushright}
\vbox{
\begin{tabular}{l}
ANL-HEP-PR-13-16
\end{tabular}
}
\end{flushright}
\vspace{0.1cm}


\title{\bf Reducing theoretical uncertainties for exclusive Higgs plus one-jet production at the LHC}

\vspace*{1cm}

\author{Xiaohui Liu}
\email[]{xiaohui.liu@northwestern.edu}
\affiliation{High Energy Physics Division, Argonne National Laboratory, Argonne, IL 60439, USA} 
\affiliation{Department of Physics \& Astronomy, Northwestern University, Evanston, IL 60208, USA}
\author{Frank Petriello}
\email[]{f-petriello@northwestern.edu}
\affiliation{High Energy Physics Division, Argonne National Laboratory, Argonne, IL 60439, USA} 
\affiliation{Department of Physics \& Astronomy, Northwestern University, Evanston, IL 60208, USA}


  \vspace*{0.3cm}

\begin{abstract}
  \vspace{0.5cm}
    
    We resum a class of large Sudakov logarithms affecting Higgs boson production in the exclusive one-jet bin at the LHC.  We extend previous results 
by calculating the full one-loop soft function for this process, which extends the accuracy of the resummation to include the leading three logarithmic corrections at each order in the QCD coupling constant.  We match this result to the next-to-leading order cross section and present a detailed numerical study assuming realistic LHC cuts.  Careful attention is paid to the matching procedure, and to the theoretical uncertainties induced by residual scale variation.  We find that the matched $\text{NLL}^{\prime}+\text{NLO}$ cross section has significantly smaller uncertainties than the fixed-order result, and can be used to alleviate the theoretical errors hindering current Higgs analyses at the LHC.

\end{abstract}

\maketitle

\section{Introduction}
\label{sec:intro}

The discovery last year of a new boson by the ATLAS and CMS collaborations at the LHC~\cite{:2012gk,:2012gu} has ushered in a new era in particle physics.  The future program of the LHC, and the next stage of experimental studies in high energy physics, will be largely devoted to measuring and understanding the properties of the new state in order to determine the underlying theory from which it arises.  The initial data provides only a hazy glimpse at the properties of the new particle.  Initial measurements of its branching ratios into various final states indicates that its couplings are consistent with those predicted for the Standard-Model Higgs boson~\cite{couplings-atlas}, as is its parity~\cite{couplings-cms}.   Significant work will clearly be needed to sharpen our picture of the new state.

A major component of the quest for a better understanding of the newly-discovered state will be the improvement of theoretical predictions for Higgs-boson production and decay channels.  It is well-known that the next-to-leading order (NLO) QCD radiative corrections to the gluon-initiated $gg \to H + X$ production process are so large~\cite{Dawson:1990zj,Djouadi:1991tka,Spira:1995rr,deFlorian:1999zd,Ravindran:2002dc,Glosser:2002gm,Campbell:2006xx} that next-to-next-to-leading order (NNLO) results are required to realistically describe LHC measurements.  NNLO calculations are available for inclusive Higgs boson production~\cite{Harlander:2002wh,Anastasiou:2002yz,Ravindran:2003um} and for the fully-differential Higgs plus zero-jet cross section~\cite{Anastasiou:2004xq,Anastasiou:2005qj,Anastasiou:2007mz,Catani:2007vq,Grazzini:2008tf}.  Recently, first results for the NNLO cross section in the Higgs plus one-jet channel have become available~\cite{Boughezal:2013uia}.  As the gluon-fusion process is the primary production mode at the LHC, it has received an enormous amount of theoretical attention.  In addition to the QCD corrections through NNLO, the leading electroweak~\cite{Aglietti:2004nj,Actis:2008ug} and mixed QCD-electroweak~\cite{Anastasiou:2008tj} corrections are known.  More complete recent reviews of precision predictions for the gluon-fusion and other production channels can be found in Refs.~\cite{Boughezal:2009fw,Dittmaier:2011ti,Dittmaier:2012vm,Grazzini:2012mi}.

Most of the theoretical predictions described above are obtained using fixed-order perturbation theory, and assume that there are no severe cuts on the phase space of the hadronic radiation produced in association with the Higgs.  Unfortunately, such constraints are present in several important Higgs search channels.   A well-known example is that of a Higgs boson decaying to $W$-bosons~\cite{Aad:2012npa,Chatrchyan:2012ty}. The background composition to this signal changes as a function of jet multiplicity.  In the zero-jet bin the background is dominated by continuum $WW$ production, while in the one-jet and two-jet bins, top-pair production becomes increasingly important.  The optimization of this search requires cuts dependent on the number of jets observed, and therefore also on theoretical predictions for exclusive jet multiplicities.  Theoretical predictions for exclusive jet bins suffer from large logarithms of the form $L=\text{ln} (Q/p_T^{veto})$, where $Q \sim M_H$ denotes the hard scale in the process.  For disparate scales $Q$ and $p_T^{veto}$, these 
logarithms can overcome the $\alpha_s$ suppression that occurs at each order in perturbation theory, and fixed-order results can consequently lead to incorrect conclusions.  For example, for the experimentally relevant values $p_T^{veto} \sim 25-30$ GeV,  residual scale variations in fixed-order calculations lead to estimated errors that do not accurately reflect uncalculated higher-order corrections~\cite{Anastasiou:2008ik,Stewart:2011cf,Banfi:2012yh}.  The importance of 
controlling these large logarithms in order to obtain reliable central values and uncertainties for the gluon-fusion channel 
in exclusive jet bins has been emphasized in the literature~\cite{Berger:2010xi,Stewart:2011cf,Gangal:2013nxa}.

The theoretical community has invested significant recent effort in attempting to resum jet-veto logarithms to all orders in perturbation theory in order to more accurately model the LHC Higgs signal.  The resummation for the zero-jet bin cross section in the presence of the anti-$k_T$ algorithm was first obtained at next-to-leading logarithmic (NLL) accuracy~\cite{Banfi:2012yh} and later extended to NNLL accuracy~\cite{Becher:2012qa,Banfi:2012jm} using two different theoretical approaches.  The importance of 
potentially large $ \text{ln} \,R$ corrections on numerical predictions, where $R$ is the jet-radius parameter in the anti-$k_T$ algorithm, was studied in Ref.~\cite{Tackmann:2012bt}.  A significant reduction of the residual theoretical uncertainties 
was obtained in the zero-jet bin by resumming the jet-veto logarithms.  Given that the theoretical uncertainties are currently 
one of the largest systematic errors affecting the one-jet bin analyses of the Higgs-like particle properties~\cite{jianming}, it is desirable to formulate the resummation when final-state jets are also present.

In a previous paper we established the formalism necessary for resummation of the Higgs plus jet process by 
deriving a factorization theorem using soft-collinear effective theory (SCET)~\cite{Bauer:2000ew,Bauer:2000yr,Bauer:2001ct,Bauer:2001yt,Bauer:2002nz} for the production of color-neutral particle and one or more jets in the presence of a 
jet-veto~\cite{Liu:2012sz}.  This result assumes that the transverse momenta of all hard jets are larger than the veto scale.  We calculated contributions through next-to-leading order in the exponent of the Sudakov form factor, and presented initial numerical 
results for Higgs production in association with a jet at the LHC.  We found that resummation of the jet-veto logarithms significantly improves the reliability of the perturbative expansion, and could potentially lead to a reduced theoretical systematic error in experimental studies.

In this manuscript we extend the calculation of Ref.~\cite{Liu:2012sz} in several ways.  We first present a calculation of the NLO soft function appearing in the factorization theorem for exclusive Higgs plus one-jet production.  This allows the extension of the resummation accuracy to $\text{NLL}^{\prime}$ accuracy, using the logarithmic counting established in Ref.~\cite{Berger:2010xi}.  This level of logarithmic accuracy implies that we correctly obtain the first three logarithmic corrections at each order in the QCD coupling constant:  $\alpha_s L^2$, $\alpha_s L$ and $\alpha_s$; $\alpha_s^2 L^4$, $\alpha_s^2 L^3$, and $\alpha_s^2 L^2$; $\alpha_s^3 L^6$, $\alpha_s^3 L^5$, $\alpha_s^3 L^4$; and so on.  We match our results to fixed-order to obtain a $\text{NLL}^{\prime}+\text{NLO}$ prediction, and present numerical results for use in LHC analyses.  We first demonstrate that the region of phase space where the leading-jet transverse momentum is of order the Higgs mass accounts for nearly half of the error in the fixed-order NLO prediction for Higgs plus one jet, and is therefore a prime candidate for an improved theoretical treatment.  We then perform a detailed study of the residual theoretical uncertainties using our resummed prediction that accounts for the variation of all unphysical scales remaining in the prediction.  Even with a very conservative treatment of the errors, a significant reduction of the residual uncertainty as compared to the fixed-order estimate is found; the estimated uncertainties decrease by up to a quarter of their initial values.  Our results, and the improvements in the zero-jet bin obtained previously, should form the basis for future theoretical error estimates in experimental analyses of Higgs properties.

Our paper is organized as follows.  We review the factorization theorem of Ref.~\cite{Liu:2012sz} in Section~\ref{secfact}.  We discuss the extension of the resummation to the $\text{NLL}^{\prime}$ level in Section~\ref{sec:extension}, and present the calculation of the previously unknown one-loop soft function.  A detailed discussion of numerical results for the LHC is given in Section~\ref{sec:num}.  We describe there how we estimate theoretical uncertainties in both the fixed-order and resummed results, and demonstrate that the resummation of jet-veto logarithms reduces the theoretical systematic error affecting LHC analyses.  Finally, we conclude in Section~\ref{sec:conc}.  Many technical details needed for the numerical studies are given in the Appendix.

\section{Review of the factorization theorem}\label{secfact}

We begin by reviewing the salient features of the factorization theorem for exclusive Higgs plus one-jet production~\cite{Liu:2012sz}.  The factorization of the cross section into separate hard, soft, and collinear sectors is complicated by the 
presence of the jet algorithm needed to obtain an infrared-safe observable.  Following the experimental analyses, 
we use the anti-$k_T$ algorithm~\cite{Cacciari:2008gp} to define jets.  Anti-$k_T$ jets are built using the following distance metrics:
\begin{eqnarray}
\rho_{ij} &=& {\rm min}(p_{T,i}^{-1},p_{T,j}^{-1})\Delta R_{ij}/R, \nonumber \\
\rho_i &=& p_{T,i}^{-1}. 
\label{metrics}
\end{eqnarray}
The anti-$k_T$ algorithm merges particles $i$ and $j$ to form a new particle by adding their four-momenta
if $\rho_{ij}$ is the smallest among all the metrics.  Otherwise, $i$ or $j$ is promoted to a jet depending on whether $\rho_i$ or $\rho_j$ is smaller, and removed from the set of considered particles.  This procedure is repeated until all particles are grouped into jets.  We note that $\Delta R_{ij}^2 = \Delta \eta_{ij}^2 + \Delta \phi_{ij}^2$,
where $\Delta \eta_{ij}$
and $\Delta \phi_{ij}$ are the rapidity and azimuthal angle
difference between particles $i$ and $j$, respectively. $R$ is the
jet-radius parameter, which in practice is chosen to be around $0.4-0.5$. 

We demand that the final state contain only a single jet with $p_T^J > p_T^{veto} \sim 25-30$ GeV.  Other jets with a transverse momentum above this threshold are vetoed. Since $p_T^{veto}$ is usually substantially lower than the partonic center-of-mass energy ($\lambda \equiv p_T^{veto}/\sqrt{\hat{s}} \ll 1$), the vetoed observables are usually very sensitive to soft and collinear emissions.  We will make the following assumptions in order to proceed in our analysis:
\begin{equation}
p_T^J \sim m_H \sim \sqrt{\hat{s}}; \;\;\; 1\gg R^2\gg \lambda^2 ; \;\;\; \frac{\alpha_s}{2\pi} \log^2 R \ll 1.
\end{equation}
The first assumption leaves us with a two-scale problem and allows the measured final-state jet to be described by a separate collinear sector.  The second of these requirements is necessary to insure that the measurement function factorizes into separate measurements in each of the collinear sectors.  The third requirement ensures that logarithms associated with the anti-$k_T$ parameter $R$ need not be resummed.  We will see later that the first assumption is satisfied in approximately $30\%$ of the relevant 
phase space for Higgs plus jet production at the LHC, and that this parameter region contributes roughly half of the total error.  We will therefore be able to improve the theoretical description of a significant fraction of the LHC Higgs signal.  Given that $p_T^{veto} \approx 25-30\,{\rm GeV}$ and $R \approx 0.4-0.5$ , when the leading jet $p_T^J \sim m_H$, the second two assumptions are also justified.

Our effective theory consists of the following low-energy degrees of freedom:
\begin{itemize}

\item a collinear jet mode with momentum 
$p_J = \frac{\w_J}{2} n_J + k_J$, where $n_J$ is the light-cone vector along the jet direction;

\item two collinear modes propagating along the beam
axes $a$ and $b$, with $p_i = \frac{\w_i}{2}n_i + k_i$ for $i = a,b$;

\item a soft mode with momentum $k_s$.

\end{itemize}
The residual momenta $k_J$, $k_i$
and the soft momentum $k_s$ all scale as $\sqrt{\hat{s}} \lambda$, while
 the large components of the three collinear momenta scale as $\w_i \sim \sqrt{\hat{s}}$.  Momenta with smaller scalings, 
 such as ultrasoft modes, do not contribute to the final-state observable and can be integrated over, and therefore need 
 not be introduced.  We are able to utilize an effective-theory framework because of how the 
 anti-$k_T$ algorithm clusters the soft and collinear modes.  Referring to the metrics defined in Eq.~(\ref{metrics}), we 
 find
 \bea
&&\rho_{JJ} \lesssim \rho_J\sim  1\,, 
\hspace{3.ex}
\rho_{Js} \sim R^{-1}\,,
\hspace{3.ex}
\rho_{Ja} \sim \rho_{Jb} \sim R^{-1}\log \lambda^{-1}\,, \nn\\
&&\rho_{ss} \sim \rho_{aa}\sim \,
\rho_{bb} \sim (\lambda R)^{-1}\,,
\hspace{3.ex}
\rho_{sa}\sim \rho_{sb} \sim \rho_{ab} \sim  (\lambda R)^{-1} \log\lambda^{-1} \,, \nn\\
&& \rho_s \sim \rho_a \sim \rho_b \sim \lambda^{-1} \,.
\eea
From $\rho_{JJ}$ and $\rho_J$ in the first line, we see that the initial clustering combines the final-state hard emissions 
into a jet, so that the soft radiation sees only the jet direction and does not probe its internal structure. Also, since the clustering
between the soft and jet radiation typically occurs earlier than
the clustering among the soft radiation, clustering of soft particles across the jet boundary is unlikely to happen~\cite{Cacciari:2008gp,Kelley:2012kj}. We see from the second line that the mixing between the soft and beam sectors is power-suppressed, as is the mixing between the beam and jet sectors.  Denoting the measurement function that imposes the jet clustering and vetoing as $\hat{{\cal M}}$, these 
factors imply that we can factor $\hat{{\cal M}}$ into the product of measurement functions acting separately on the 
soft, jet, and beam sectors,
\bea
\hat{\cal M} = \hat{\cal M}_J\hat{\cal M}_s\hat{\cal M}_a\hat{\cal M}_b,
\eea
up to power-suppressed corrections in $p_T^{veto}$ and $R$.

The remaining steps in the derivation of the factorization theorem utilize the standard SCET machinery, and are presented in detail in Ref.~\cite{Liu:2012sz}.  The final result for the cross section for exclusive Higgs plus one-jet production 
takes the following form:
\bea\label{factgen}
\mathrm{d}\sigma_{\text{NLL}^{\prime}} &=& \mathrm{d}\Phi_{H}\mathrm{d}\Phi_{J}\,
{\cal F}(\Phi_{H},\Phi_{J})
\,
\sum_{a,b}\int \mathrm{d}x_{a} \mathrm{d}x_b \frac{1}{2\hat{s}}\,
 (2\pi)^4 \delta^4\left(q_a + q_b - q_{J} -q_{H}\right)\nn\\
&&\times 
\bar{\sum_{\rm spin}}
\bar{\sum_{\rm color}}
{\rm Tr}(H\cdot S)\,
{\cal I}_{a,i_aj_a} \otimes f_{j_a}(x_a)\,
{\cal I}_{b,i_bj_b} \otimes f_{j_b}(x_b)
J_{J}(R)\,.
\eea
We have denoted explicitly by the subscript that we will evaluate this cross section to the $\text{NLL}^{\prime}$ level.  $\mathrm{d}\Phi_{H}$ and $\mathrm{d}\Phi_{j_i}$
are the phase space measures for the Higgs and 
the massless jet $J$, respectively. ${\cal F}(\Phi_{H_c},\Phi_{J})$ includes all additional
phase-space cuts other than the $p_T$ veto acting on the Higgs boson and the hard jet.  $H$ is the hard function that 
comes from matching full QCD onto SCET, and $S$ describes soft final-state emissions.  The trace is over the color indices.  The functions ${\cal I}$ and $J$ describe collinear emissions along the beam axes and along the final-state jet direction, respectively.  The measured jet $p_T^J$ should be much larger than $p_T^{veto}$.  Operator definitions 
for all functions are given in Ref.~\cite{Liu:2012sz}.  As our purpose here is to only briefly review the factorization theorem before presenting new results, we do not reproduce these definitions explicitly, and instead refer the reader to the quoted reference.  The tree-level and one-loop expressions for the jet and beam functions needed for numerical studies are 
presented in the Appendix.

\section{Extension to $\text{NLL}^{\prime}$}\label{sec:extension}

A primary goal of this manuscript is to extend the resummation of jet-veto logarithms to the $\text{NLL}^{\prime}$ level, following the notation of Ref.~\cite{Berger:2010xi}.  This level of logarithmic accuracy implies that we correctly obtain the following towers of logarithms:  $\alpha_s L^2$, $\alpha_s L$ and $\alpha_s$; $\alpha_s^2 L^4$, $\alpha_s^2 L^3$, and 
$\alpha_s^2 L^2$; $\alpha_s^3 L^6$, $\alpha_s^3 L^5$, $\alpha_s^3 L^4$; and so on.  The following ingredients 
are required to obtain this accuracy:
\begin{itemize}

\item the two-loop cusp and one-loop non-cusp anomalous dimensions which control the evolution of the beam, jet, soft  and hard functions;

\item the one-loop hard, beam and jet functions;

\item the one-loop soft function.

\end{itemize}
The requisite anomalous dimensions, as well as the one-loop jet and beam functions, were obtained in Ref.~\cite{Liu:2012sz}.  They are included in the Appendix for completeness, as is a detailed discussion of their implementation into Eq.~(\ref{factgen}).  The one-loop hard function for the $gg$, $qg$ and $q\bar{q}$ channels can be obtained from the literature~\cite{Schmidt:1997wr}.  The previously unknown quantity is the one-loop soft function.  Its calculation is 
rendered non-trivial by the presence of the final-state jet.  We describe our computation of the soft function below.

\subsection{Calculation of the one-loop soft function}

We begin by defining the measurement function for the soft sector:
\begin{equation}
\hat{\cal M}_s = \Theta_{p_T^{veto},k_T}\Theta_{\Delta {R_{kJ}},R}\,
+ \Theta_{R,\Delta {R_{kJ}}},
\end{equation}
where we have set $\Theta_{a,b} = \theta(a-b)$ and $\Delta R_{kJ} =\sqrt{ \Delta y^2 + \Delta \phi^2}$.  The first term allows soft emissions that are well-separated from the final-state jet but have a transverse momentum softer than 
$p_T^{veto}$, while the second term allows harder emissions that are within the final-state jet radius.  As the soft function is simply the square of the soft current integrated over the allowed phase space, we can immediately write the NLO contribution to the soft function as
\bea
S &=& -\frac{2g_s^2}{(2\pi)^{d-1}} \sum_{i<j}T_i\cdot T_j  \,
\int \mathrm{d}^d k \, \delta(k^2) \frac{n_i\cdot n_j}{n_i\cdot k n_j\cdot k}\,\hat{\cal M}_s,
\eea
where the sum is over the two beam directions and the final-state jet direction.  The $n_i$ denote light-like vectors in each of these directions, while the $T_i$ denote color operators in either the fundamental or adjoint representations, depending on whether $i$ denotes a quark or a gluon.  $k$ is the momentum of the gluon emitted from the eikonal lines.   We note that the leading-order soft function is normalized to unity. Parameterizing
\bea
k^\mu &=& k_T \,(\cosh y, \cos \phi, \sin \phi, \sinh y) \,,\nn \\
n_J^\mu &=& (\cosh y_J,1, 0, \sinh y_J),
\eea
and setting
\be
\Theta_{\Delta {R_{kJ}},R} = 1-  \Theta_{R,\Delta {R_{kJ}}} \,,
\ee
we find
\bea
S &=& -g_s^2 \frac{\Omega_{1-2\epsilon}}{2 (2\pi)^{d-1}}
\sum_{i<j}T_i\cdot T_j  \,
\int \mathrm{d} y \,
\mathrm{d}\phi \,
\mathrm{d}k_T^2 \, 
(s_\phi)^{-2\epsilon}\, (k_T^2)^{-\epsilon}
 \frac{n_i\cdot n_j}{n_i\cdot k n_j\cdot k} \nn \\ 
&\times & \left(\Theta_{p_T^{veto},k_T}
+ \Theta_{R,\Delta {R_{kJ}}} \Theta_{k_T,p_T^{veto}} \, \right)\,
\label{softfunc1}
\eea
where $\Omega_d$ denotes the $d$-dimensional solid angle and $s_{\phi} = \text{sin} \, \phi$.

We will proceed by reducing these integrals as far as possible analytically, although we will end up with two remaining integrals which we evaluate numerically.  Since the first such integral is a pure number, while the second depends only upon $R$, this does not affect the speed of the numerical program we construct.  We note that the soft function will have rapidity divergences arising from the $n_i \cdot k$  in the denominator of Eq.~(\ref{softfunc1}), which we will regulate by multiplying the integrand by the following factor~\cite{Chiu:2012ir}:
\begin{equation}
|2k_{g,3}|^{-\eta}\nu^\eta.
\label{rapreg}
\end{equation}
This regulates the rapidity divergence as a pole in $\eta$, which is then removed by renormalization. There are two distinct structures to consider in Eq.~(\ref{softfunc1}): the first when both $i$ and $j$ denote a beam direction, and the second when $i$ denotes the jet direction.  We write the soft function as 
\begin{equation}
S = T_a \cdot T_b \, S_{n\bar{n}} + T_a \cdot T_J \, S_{nJ} + T_b \cdot T_J \, S_{\bar{n}J}
\label{softcolstruc}
\end{equation}
and study each structure separately.  We note before continuing that the virtual corrections are scaleless, and have 
the effect of converting the infrared poles in the real-emission corrections into ultraviolet poles.

\medskip
\noindent
\underline{$S_{n\bar{n}}$}: This case will have rapidity divergences as $n \cdot k \to 0$ and $\bar{n} \cdot k \to 0$.  We can 
therefore replace the regulator of Eq.~(\ref{rapreg}) using
\bea
|2k_{g,3}|^{-\eta}\nu^\eta \,
\xrightarrow{|y|\to \infty } k_T^{-\eta} \nu^{\eta} \exp(-\eta\, |y|)\,.
\eea
This relation is valid in the large $y$ limit in which the rapidity divergence occurs.  Since we are treating ${\cal O}(R)$ contributions as power-suppressed terms and there are no singularities associated with emissions along the final-state jet direction, we can expand $\Theta_{\Delta R_{kJ},R} = 1+{\cal O}(R)$ to derive
\be
S_{n\bn} =
g_s^2\frac{2\Omega_{2-2\epsilon}}{(2\pi)^{d-1}}  \,
\frac{1}{2\epsilon + \eta}
 \left(\frac{p_T^{veto}}{\mu}\right)^{-2\epsilon-\eta}\,
\left(\frac{\nu}{\mu}\right)^\eta\,
\frac{2}{\eta}.
\ee
This can be easily expanded in both $\eta$ and $\epsilon$ in order to isolate the poles, which are then removed by renormalization.

\medskip
\noindent
\underline{$S_{n n_J}$}: The calculation of this structure is more involved than the previous one.  We first note that 
the $n \cdot k \to 0$ rapidity divergence is correctly regulated by the following replacement, valid in the large-$y$ limit in which the divergence occurs:
\bea
|2k_{g,3}|^{-\eta}\nu^\eta = 2^{-\eta}k_T^{-\eta}\nu^\eta|\sinh y|^{-\eta}
\xrightarrow{y\to \infty } k_T^{-\eta} \nu^{\eta} \exp(-\eta\, y\,\Theta(y) )\,.
\eea
Performing the $k_T$ integration in Eq.~(\ref{softfunc1}), and afterwards using the relation $\Theta_{\Delta {R_{kJ}},R} = 1-  \Theta_{R,\Delta {R_{kJ}}}$, we find 
\bea
S_{nn_J} &=&
g_S^2\frac{\Omega_{1-2\epsilon}}{(2\pi)^{d-1}}  \,
\frac{\pi}{2\epsilon+\eta} \,
\left(\frac{p_T^{veto}}{\mu}\right)^{-2\epsilon - \eta}\,
\left(\frac{\nu\exp(-y_J)}{\mu} \right)^\eta \nn \\
&&\hspace{2.ex} \times \,
\int_{-\infty}^\infty \mathrm{d} \Delta y \,
\int_0^\pi\frac{\mathrm{d}\phi}{\pi} \, 
(s_{\phi})^{-2\epsilon}\, 
 \frac{\exp(\Delta y)\exp(-\eta \Delta y\,\Theta(\Delta y))}{(\cosh\Delta y-\cos \phi)}\,
\Theta_{\Delta {R_{kJ}},R}  \,,
\eea
where we have shifted the integration variables to $\Delta y = y-y_J$.  We can no longer directly expand the integrand for $R \to 0$, as there is a logarithmic divergence in this limit associated with collinear emissions along the final-state jet direction.  We will proceed by isolating the small-$R$ behavior of the integrand by defining
\bea
{\cal I} &\equiv& \,
(s_{\phi})^{-2\epsilon}\, 
 \frac{\exp(\Delta y)\exp(-\eta \Delta y\,\Theta(\Delta y))}{(\cosh\Delta y-\cos \phi)}\,, \nn \\
{\cal I}_R &\equiv& \,
(s_{\phi})^{-2\epsilon}\, 
 \frac{2\,\exp(-\eta \Delta y\,\Theta(\Delta y))}{(\Delta y^2+\phi^2)}\,.
\eea
We divide the integrand into the two structure ${\cal I} - {\cal I}_R$ and ${\cal I}_R$.  The first piece is finite as $R \to 0$ and can be Taylor-expanded in that limit, while the second contains $\text{ln} \, R$ behavior and cannot.  We further find it convenient for the ${\cal I} - {\cal I}_R$ term to divide the $\Delta y$ integral into two regions: $\Delta y <0$, for which the 
rapidity divergence cannot occur and consequently we can set $\eta = 0$; $\Delta y>0$, in which the rapidity 
divergence can occur and $\eta$ must be retained.  Performing the indicated manipulations, we reduce $S_{nn_J}$ into the sum of three structures:
\be
S_{nn_J} = S_{nn_J}^{\text{ln}R} + S_{nn_J}^{\eta} + S_{nn_J}^{\eta=0},
\ee
where we have identified
\bea
S_{nn_J}^{\text{ln}R} &=& g_s^2 \frac{\Omega_{1-2\epsilon}}{(2\pi)^{d-1}}  \,
\frac{\pi}{2\epsilon} \,
\left(\frac{p_T^{veto}}{\mu}\right)^{-2\epsilon} \,
\int_{-\infty}^\infty \mathrm{d} \Delta y \,
\int_0^\pi\frac{\mathrm{d}\Delta \phi}{\pi} \, 
(s_{\Delta \phi})^{-2\epsilon}\, 
 \frac{2}{\Delta {R_{kJ}}^2}\,
\Theta_{\Delta {R_{kJ}},R}  \,, \nn \\
S^\eta_{nn_J} &=& g_s^2
\frac{\Omega_{2-2\epsilon}}{(2\pi)^{d-1}}  \,
\frac{1}{2\epsilon+\eta} \,
\left(\frac{p_T^{veto}}{\mu}\right)^{-2\epsilon - \eta}\,
\left(\frac{\nu\exp(-y_J)}{\mu} \right)^\eta 
\frac{2}{\eta}\,, \nn \\
S_{nn_J}^{\eta=0} &=& g_s^2\frac{\Omega_{1-2\epsilon}}{(2\pi)^{d-1}}  \,
\frac{\pi}{2\epsilon} \,
\left(\frac{p_T^{veto}}{\mu}\right)^{-2\epsilon }\, 
\int_{-\infty}^\infty \mathrm{d} \Delta y \,
\int_0^\pi\frac{\mathrm{d}\Delta \phi}{\pi} \, 
\left[{\cal I}-{\cal I}_R - 2(s_{\Delta \phi})^{-2\epsilon}\Theta(\Delta y) \right]\,.
\label{snnj}
\eea
We will numerically calculate the integrals which are left upon expansion of the integrands in $\epsilon$.  We note that the $S_{\bn n_J}$ term of Eq.~(\ref{softcolstruc}) can be obtained by taking $y_J \to -y_J$ in Eq.~(\ref{snnj}), which affects 
only the $S^\eta_{nn_J}$ structure above.

Combining all of the information presented above, expanding in both $\eta$ and $\epsilon$ and removing the poles via renormalization, we are left with the following final result for the soft function:
\bea
S &=& \frac{\alpha_s}{4\pi} \left\{ (T_a^2+T_b^2) \left[ L^2 + 4 \, \text{ln} \frac{p_T^{veto}}{\nu} \, L\right] +2 \, T_J^2 \,L \, \text{ln} \, R^2 +4 \,  y_J \, L \left( T_a \cdot T_J - T_b \cdot T_J \right)\right. \nn \\
&-& (T_a^2+T_b^2) \frac{\pi^2}{6} + T_J^2 \left[ c +f(R)\right]\,,
\label{finalsoft}
\eea
where we have abbreviated $L = \text{ln} (\mu^2 / p_T^{veto,2})$.  The constant $c$ and function $f(R)$ 
are given by the following integrals:
\bea
c &=& 4  \,
\int_{-\infty}^\infty \mathrm{d} \Delta y \,
\int_0^\pi\frac{\mathrm{d}\phi}{\pi} \, 
\left({\cal I}^{\eta=0,\epsilon=0}-{\cal I}^{\eta=0,\epsilon=0}_R
- 2\Theta(\Delta y)\right)\log(s_{\phi})\,, \nn \\
f(R) &=& -4\log(2) \log R^2 \,
+8\int_{-\infty}^\infty \mathrm{d} \Delta y \,
\int_0^\pi\frac{\mathrm{d}\phi}{\pi} \, 
\frac{\log(s_{\phi})}{\Delta {R_{kJ}}^2}\,
\Theta_{\Delta {R_{kJ}},R}  \,.
\eea
As stated above, we determine these quantities numerically in our analysis.  As we only perform our analysis for a very small set of $R$ values, this is sufficient for our purposes.  We note that the $L^2$ and $L$ terms agree with those predicted in 
Ref.~\cite{Liu:2012sz} using the cancellation of the combined running of the hard, jet, beam and soft functions.
For the Higgs production process considered in this 
work, we have the following color identities: for the $ggg$ channel, $T_i\cdot T_j = -C_A/2$; for the 
$q_1{\bar q}_2g_3$ channel, 
$T_1\cdot T_2 = -(C_F-C_A/2)$ and $T_1\cdot T_3 = T_2\cdot T_3 = -C_A/2$.

\section{Numerical results}\label{sec:num}

We now present numerical results for exclusive Higgs plus one-jet production at the LHC.  We first discuss the matching of the resummed result with the NLO cross section to obtain a renormalization-group (RG) improved $\text{NLL}^{\prime}+\text{NLO}$ prediction, and demonstrate that we correctly capture the large logarithms associated with $p_T^{veto}$.  We also discuss the parameter region in which our effective-theory framework is valid.  Although this region, with $p_T^J \sim m_H$, makes up only $\sim$25-30\% of the signal, it accounts for nearly half of the theoretical uncertainty in the one-jet bin.  We then describe in detail how we estimate the theoretical uncertainties in both the fixed-order and RG-improved results.  For the fixed-order cross section we follow the ``ST" recommendations of Ref.~\cite{Stewart:2011cf}.  Our treatment of the theoretical uncertainties of the RG-improved result is necessarily more involved, as our effective-theory approach improves the prediction over only part of the relevant phase space.  We adopt a combination of direct scale variation, which is standard in resummed calculation~\cite{Stewart:2011cf}, and the ST recommendations for the fixed-order region, as described below.  Finally, we show that the resummation of the jet-veto logarithms leads to a sizable reduction of the exclusive Higgs plus one-jet uncertainties at the LHC.

\subsection{Matching $\text{NLL}^{\prime} $ with NLO}

We begin by matching our resummed expression with the fixed-order NLO result to obtain a $\text{NLL}^{\prime}+\text{NLO}$ prediction.  We use the NLO predictions for Higgs plus one-jet contained in MCFM~\cite{Campbell:2010ff}.  We obtain our prediction by setting
\bea\label{rgimproved}
\sigma_{\text{NLL}^{\prime}+\text{NLO}} = \sigma_{{\rm NLL}'} + \,
\sigma_{\rm NLO}-\sigma^{\rm exp}_{{\rm NLL}'}.
\eea
In this equation, $\sigma_{\rm NLO}$ is the fixed-order NLO 
cross section obtained from MCFM, and $\sigma_{\rm NLL}^{\prime}$ is the resummed 
cross section up to $\text{NLL}^{\prime}$ accuracy presented in Eq.~(\ref{factgen}).  $\sigma^{\rm exp}_{\rm NLL'}$
captures the singular features of $\sigma_{\rm NLO}$, and is obtained by expanding 
$\sigma_{\rm NLL'}$ with all scales set to a common value $\mu$.  Schematically, we have 
\bea
&&\sigma_{{\rm NLL}'} = \,
\sigma_{\rm LO} \left(1 + \alpha_s  g_0\right)
 e^{-L g_{\rm LL}(\alpha_s L)-g_{\rm NLL}(\alpha_s L)}\, \nn \\
&&\sigma^{\rm exp}_{{\rm NLL}'} = \,
\sigma_{\rm LO}\left(1 + \alpha_s \left[ -g_2 L^2 - g_1 L \,
+ g_0\right] \right)\,,
\eea
where $L\,g_{\rm LL}$ and $g_{\rm NLL}$ resum the leading and next-to-leading logarithms,
respectively.   The difference between
$\sigma_{\rm NLO}$ and the
expanded NLL' result $\sigma^{\rm exp}_{{\rm NLL}'}$ only
contains power-suppressed contributions for large values of $Q$:
\bea
\sigma_{\rm non-singular}\equiv \sigma_{\rm NLO}-\sigma^{\rm exp}_{{\rm NLL}'}
\sim {\cal O}\left(R^2 L,\frac{p_T^{veto}}{Q}L\,,
\frac{p_T^{veto}}{Q}\log R \,,
\cdots \right) \,,
\eea
with $L = \log \left(Q/p_T^{veto}\right)$, and $Q$ stands for any kinematic quantity of order $m_H$. Since
the scale $Q R$ is used to define the jet mode, the $R^2L$ terms are regarded as power suppressed.

\begin{figure}[h!]
\begin{center}
  \includegraphics[width=3.7in,angle=0]{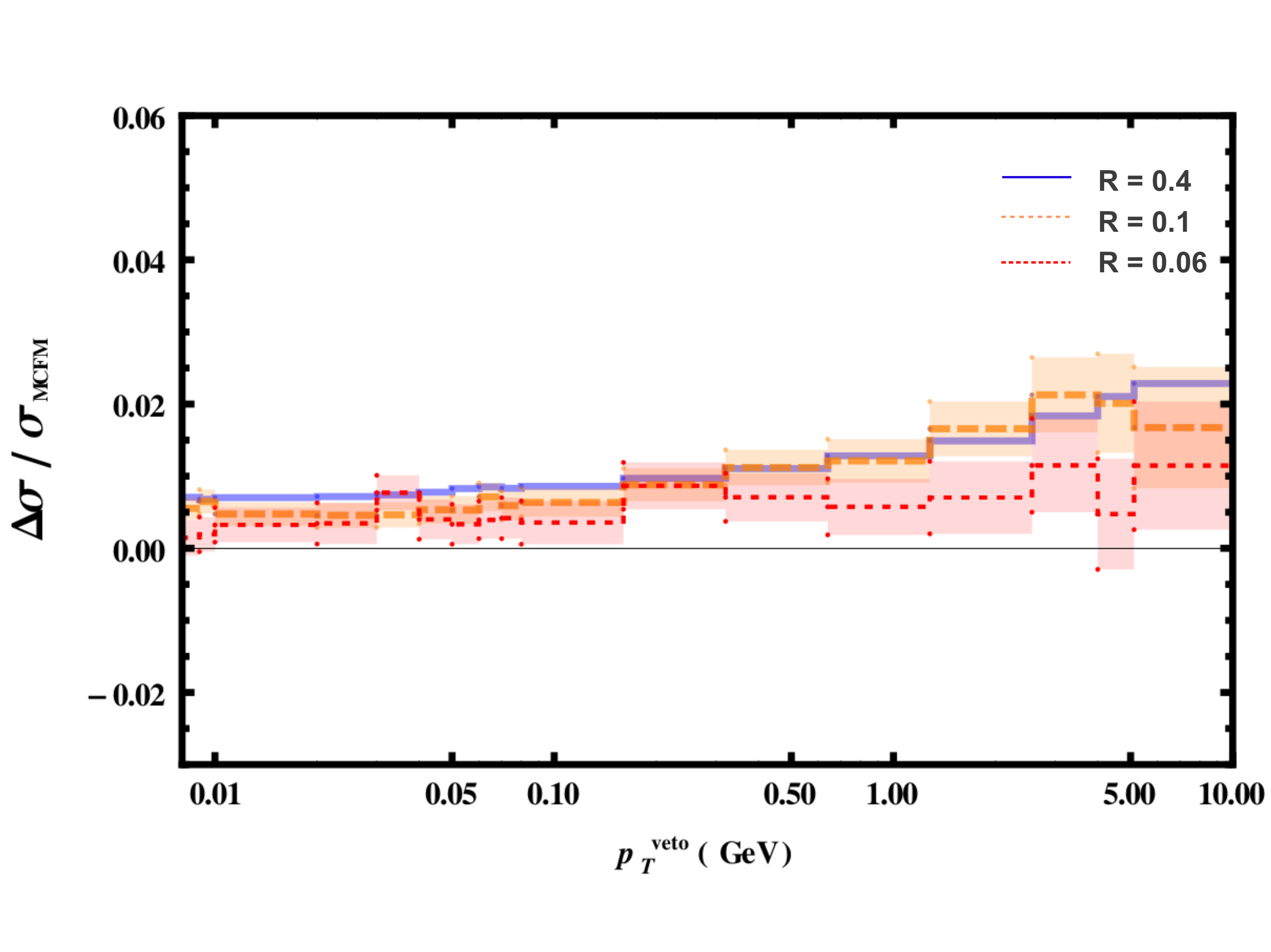}
\end{center}
\vspace{-0.1cm}
\caption{Shown are the fractional differences $(\sigma_{\rm NLO}-\sigma^{\rm exp}_{{\rm NLL}'})/\sigma_{\rm NLO}$ for $R = 0.4$ (solid blue), $0.1$(dashed orange) and $0.06$ (dotted red), respectively. The colored bands represent the estimated numerical uncertainties. The differences between the expanded $\text{NLL}^{\prime}$ and the fixed-order NLO calculations
are small compared with the total cross section.}
\label{mcfmexpcom}
\end{figure}

To demonstrate that our formalism correctly captures the singular terms at NLO as $L \to 0$, we plot in Fig.~\ref{mcfmexpcom} the fractional difference between the expanded cross section $\sigma^{\rm exp}_{\rm NLL'}$ and the NLO result as a function of $p_T^{veto}$.  We note that due to the power-suppressed $R^nL$ terms, the difference
does not completely vanish as $p_T^{veto}$ goes to $0$ for fixed $R$.  However, we see from the several $R$ values plotted in Fig.~\ref{mcfmexpcom} that in the $R \to 0$ limit, these terms also vanish, as expected. 

\begin{figure}[!ht]
\begin{center}
  \includegraphics[width=3.7in,angle=0]{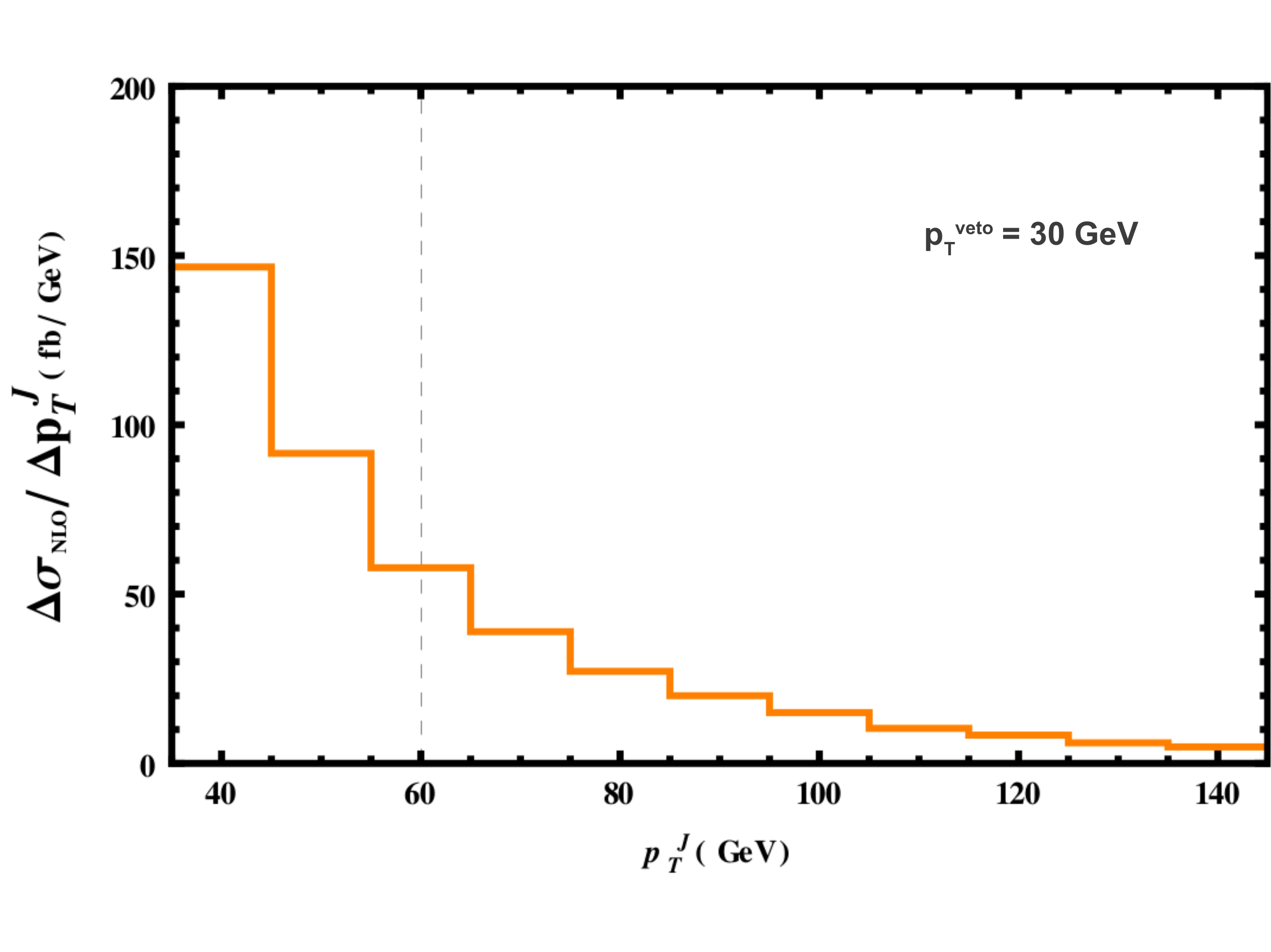}
\end{center}
\vspace{-0.5cm}
\caption{$\Delta\sigma/\Delta p_T^J$ is the bin-integrated cross section for Higgs plus one jet as a function of $p_T^J$ divided by the bin width.  We assume $p_T^{veto} = 30\,{\rm GeV}$.  If we define the lower boundary of the $p_T^J \sim {\cal O}(m_H)$ region by $p_T^J=m_H/2$, we can estimate the contribution from $p_T^J \sim {\cal O}(m_H)$ to be around $30\%$. }\label{NLOsigma}
\end{figure}

\subsection{Validity of the effective theory}
\label{sec:valid}

Having demonstrated that our $\text{NLL}^{\prime}$ result correctly captures the logarithms of $p_T^{veto}$, we comment here briefly on its expected range of validity.  In our derivation of the factorization theorem, we assumed that the signal jet $p_T^J$ is of
order $m_H$. From Fig.~\ref{NLOsigma}, we see that this configuration contributes
a non-negligible fraction of the 
experimentally-interesting
total cross section for $p_T^{veto} \sim 30$ GeV and $p_T^J > p_T^{veto}$. 
Our factorization theorem holds for $p_T^{veto}\ll p_T^J \sim Q $, but breaks down when 
$p_T^J \sim p_T^{veto} \ll m_H$.  Additional large logarithms of the form $\text{ln}^2 m_H/p_T^J$ and $L \times \, p_T^{veto}/p_T^J$ are not resummed in our formalism.  We describe these terms only as well as a fixed-order NLO calculation.  A different effective theory is needed for this regime to correctly sum the large logarithms.  We do not consider 
this theory in this manuscript; our goal here is to consistently apply the currently available formalism at $\text{NLL}^{\prime}+\text{NLO}$ to see to what extent we can reduce the theoretical uncertainty.  

Interestingly, the $p_T^J \sim m_H$ region contributes roughly 50\% of the uncertainty in the one-jet bin, larger than might be expected from Fig.~\ref{NLOsigma}.  We show this by computing the NLO cross section for an example parameter choice.  We set $m_H=126$ GeV and $p_T^{veto}=25$ GeV, and divide the Higgs plus one-jet cross section, whose inclusive value is $\sigma^{1j}_{\text{NLO}} = 5.75^{+2.03}_{-2.66}\,\text{pb}$, 
into two bins: the first with $p_T^J < m_H/2$, and the second with 
$p_T^J > m_H/2$.  As explained in detail later in Sec.~\ref{sec:uncert}, we use the fixed-order cross section in the first bin since our effective-theory analysis does not hold, and turn on resummation in the second bin.  Computing the cross section at NLO in each bin, and estimating the uncertainties as described in detail in Sec.~\ref{sec:uncert}, we find
\bea
\sigma^{1j}_{\text{NLO}} (p_T^J < m_H/2) &=& 4.74^{+1.31}_{-1.29} \, \text{pb}, \nn \\
\sigma^{1j}_{\text{NLO}} (p_T^J > m_H/2) &=& 1.01^{+0.85}_ {-1.51} \, \text{pb}.
\eea
The central values have been obtained using the scale choice for $\mu=m_H/2$.\footnote{We note that using a larger central scale choice leads to the same conclusions regarding the relative uncertainties of the two bins.}  Although it accounts for less than 25\% of the cross section, the region where our effective-theory analysis can improve the uncertainties contributes roughly half of the error in the full one-jet bin. 

We briefly comment here on non-global logarithms~\cite{Dasgupta:2001sh} that first occur at the $\text{NLL}^{\prime}$ level.  Although they are not included in our current factorization theorem, to estimate their numerical effect we use the large-$N_c$ resummation of these terms derived in Ref.~\cite{Dasgupta:2001sh}.  We include them as a multiplicative correction to our factorization formula.  Their numerical effect is small, at or below one percent of the total exclusive Higgs plus one-jet production rate for the relevant values of $m_H$ and $p_T^{veto}$.  To check the robustness of this result we vary the hard scale appearing in these corrections by a factor of two around their nominal value of $m_H$, and find similarly small corrections.  We therefore believe that it is numerically safe to neglect these terms in our $\text{NLL}^{\prime}$ result, although they should be further investigated in the future.

\subsection{Scale choices and uncertainty estimation}
\label{sec:uncert}

\begin{figure}[!ht]
\begin{center}
  \includegraphics[width=3.7in,angle=0]{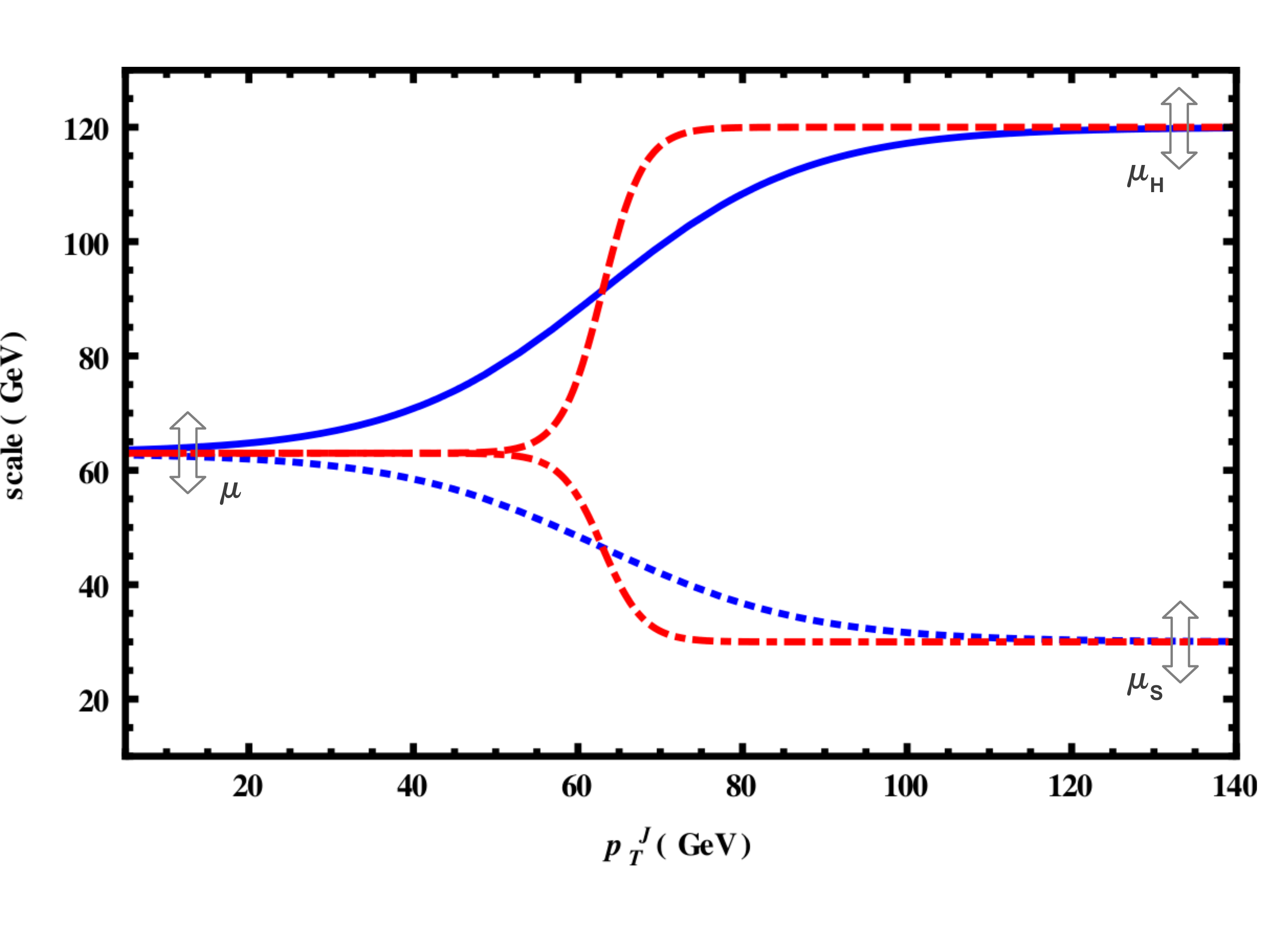}
\end{center}
\vspace{-0.5cm}
\caption{Shown is the interpolation between the resummed result and the fixed-order one proposed in Eq.~(\ref{inter}). The blue solid and dotted lines are for $\kappa = 0.04$ and the red dashed and dot-dashed lines 
are for $\kappa = 0.2$.   When $p_T^J$ is less
than $p_{\rm off}$, all the scales merge to $\mu$ and the cross section takes its NLO value.  We have demonstrated this behavior for the hard and soft scales in the plot. }\label{scalesonoff}
\end{figure}

Since the resummation holds only for $p_T \sim m_H$, we wish to turn it off and recover the fixed-order NLO result as 
$p_T^J$ becomes small.   To do so, we note that the fixed order cross section $\sigma_{\rm NLO}$ 
and the expanded $\text{NLL}^{\prime}$ cross section 
$\sigma^{\rm exp}_{{\rm NLL}'}$ depend only on the scale $\mu_R = \mu_F = \mu$, while $\sigma_{{\rm NLL}'} $ also depends on the scales
$\mu_H$, $\mu_J$, $\mu_B$, $\mu_S$, $\nu_B$ and $\nu_S$ that appear in the hard, jet, beam and soft functions. 
The optimal choice for each scale can be determined by minimizing the higher order corrections
to each separate component.  These functions are then RG-evolved to the common scale $\mu$.  Consequently, the resummation can be turned off by setting all scales
to $\mu$ , so that the full $\text{NLL}^{\prime}+\text{NLO}$ result reduces to the NLO one.  We adopt a conservative scheme to turn off the resummation as soon as possible,
as suggested in Ref.~\cite{Berger:2010xi}.  In the region where
$p_T^J \gg p_T^{veto}$, we keep the resummation on.   When $p_T^J \sim p_T^{veto}$, we switch off the resummation by setting all scales to $\mu$, which leads to the fixed-order prediction. We interpolate between these two regions smoothly using
\bea\label{inter}
\mu_i^{int.} = \mu + 
(\mu_i - \mu)\left[ \,
1+\tanh\left( \kappa\left(p_T^J-p_{\rm off}\right) \right)\,
 \right]/2\,,
\eea
where the index $i$ runs over all appearing scales.  We use similar expressions for the $\nu$'s.  Our numerical predictions 
are obtained using the $\mu_i^{int.}$ expressions in our code.   When $p_T^J < p_{\rm off}$, the resummation starts to vanish. We set 
$p_{\rm off} = \max(2p_T^{veto},\frac{m_H}{2})$ to be the default value.\footnote{The reason for this choice is that our EFT is valid when 
$p_T^J$ is located in the hard domain whose lower boundary is 
estimated to be $\frac{m_H}{2}$, and it entirely breaks down
when $p_T^J$ falls into the ``soft" regime whose upper boundary is 
roughly $2p_T^{veto}$.}
The slope $\kappa$ controls how smoothly we turn off the resummation. We find that the interpolated cross section is insensitive to the choice of $\kappa$. The functional forms of the interpolation in Eq.~(\ref{inter}) for various values $\kappa$ are shown in Fig.~\ref{scalesonoff}.  When making uncertainty estimations, we vary each scale separately. In the resummation region, the cross section is relatively insensitive to the variation of $\mu$.  In the fixed-order range, it is insensitive to $\mu_i$ and $\nu_i$. 

We describe here in detail how we estimate the uncertainties in both the fixed-order and RG-improved results.  We vary all  scales appearing in the cross section around their central values by factors of two in both directions in order to estimate the theoretical error. To avoid an underestimate of the uncertainty of the fixed-order calculation, we follow the procedure suggested by Stewart and Tackmann~\cite{Stewart:2011cf}.  We split the exclusive one-jet cross section into the difference of one-jet inclusive and two-jet inclusive results:
\bea
\sigma_{1j} = \sigma_{\ge 1j} - \sigma_{\ge 2j} \,.
\eea 
We estimate the scale uncertainty for each piece separately and 
add them in quadrature to obtain the scale uncertainty for the 
exclusive cross section:
\bea\label{UNLO}
\delta^2_{1j,{\rm NLO}} = \delta^2_{\ge 1j,{\rm NLO}} \,
 + \delta^2_{\ge 2j,{\rm NLO}} \,.
\eea
For the $\text{NLL}^{\prime}+\text{NLO}$ result, the uncertainty is derived by adding in quadrature the separate variations of all scales which enter~\cite{Stewart:2011cf}: 
\bea\label{UNLL}
\delta^2_{1j} = \delta^2_{{\rm non-singular},\mu} \,
 + \delta^2_{{\rm NLL'},\mu} \,
 + \delta^2_{{\rm NLL'},\mu_H} \,
 + \delta^2_{{\rm NLL'},\mu_J} \,
 + \delta^2_{{\rm NLL'},\mu_B,\nu_B} \,
 + \delta^2_{{\rm NLL'},\mu_S,\nu_S} \,.
\eea
Before continuing we comment briefly on the structure of Eq.~(\ref{UNLL}).  In order to perform the matching to 
fixed order in Eq.~(\ref{rgimproved}), we RG-evolve the $\text{NLL}^{\prime}$ result so that all scales are set to the common scale $\mu$.  We then add on the non-singular NLO terms via the difference between the full NLO cross section and the expanded $\text{NLL}^{\prime}$ results.  This explains the first two contributions to the above equations.  As the hard and jet functions live at the scales 
$\sqrt{m_H p_T^J}$ and $p_T^J R$ respectively~\cite{Liu:2012sz}, these scale variations are treated as uncorrelated in 
Eq.~(\ref{UNLL}).  Finally, the variations of beam and soft functions, which live at the scale $p_T^{veto}$, are added to this.

When we apply this formalism and assume actual LHC kinematic cuts, a large fraction of the cross section comes from
the low-$p_T^J$ regime where $p_T^J < p_{\rm off}$, and the fixed-order 
calculation dominates. In this situation, we split the 
cross section into two regions, one with
$p_T^{veto}<p_T^J < p_{\rm off}$ and
the other with $p_T^J > p_{\rm off}$. For the former region, we use
Eq.~(\ref{UNLO}) to estimate the uncertainty and for the latter one,
we utilize Eq.~(\ref{UNLL}). We combine these two linearly to 
estimate the scale dependence for the RG-improved cross section:
\bea\label{UMIX}
\delta_{1j}(p^J_T>p_T^{veto}) = \,
\delta_{1j,{\rm NLO}}(p^J_T<p_{\rm off}) \,
+ \delta_{1j}(p^J_T>p_{\rm off}) \,.
\eea
Since the resummation in the result used for $p_T^J > p_{\rm off}$ is turned off quickly by using the interpolation in Eq.~(\ref{inter}), and the uncertainty of the fixed-order cross section used for $p_T^J < p_{\rm off}$ is obtained using the ST prescription, we believe that this leads to a very conservative estimate of the theoretical error after performing our 
RG-improvement.

\subsection{Numerics for the LHC}

We now present predictions and uncertainty estimates for use in LHC analyses.  For the following numerical results, and those shown above, we use the MSTW 2008 parton distribution functions~\cite{Martin:2009bu} at NLO.  We assume an 8 TeV LHC, and $m_H=126$ GeV unless stated otherwise.  We demand that the leading jet be produced with rapidity $|y_J|< 4.5$, and veto all other jets with $p_T > p_T^{veto}$ over the entire rapidity range.  The following central values are used for the scales which appear:
\bea\label{cscale}
&&\mu = \sqrt{(m^T_H)_{\rm min} (p_{T}^J)_{\rm min}}, 
\hspace{5.ex} \mu_H = \sqrt{m^T_H p_T^J} \,,\nn \\
&&\mu_J = p_T^J R \,,
\hspace{5.ex} \mu_B = \mu_S = p_T^{veto} \,, \nn \\
&&\nu_{B_{a,b}}  = x_{a,b}\sqrt{ s}\,,
\hspace{5.ex} \nu_S = p_T^{veto}\,.
\eea
where $m^T_H = \sqrt{m_H^2+p_T^{J,2}}$.  We note that these central scale values, as well as the variations up and down by a factor of two, are used as the $\mu_i$ on the right-hand side of Eq.~(\ref{inter}).  The actual numerical scale choices used in the code are the $\mu_i^{int.}$ appearing on the left-hand side of Eq.~(\ref{inter}).  We use $\kappa=0.2$ to 
produce all numerical results, although we have checked that their dependence on $\kappa$ is negligible.

\begin{figure}[!ht]
\begin{center}
  \includegraphics[width=3.7in,angle=0]{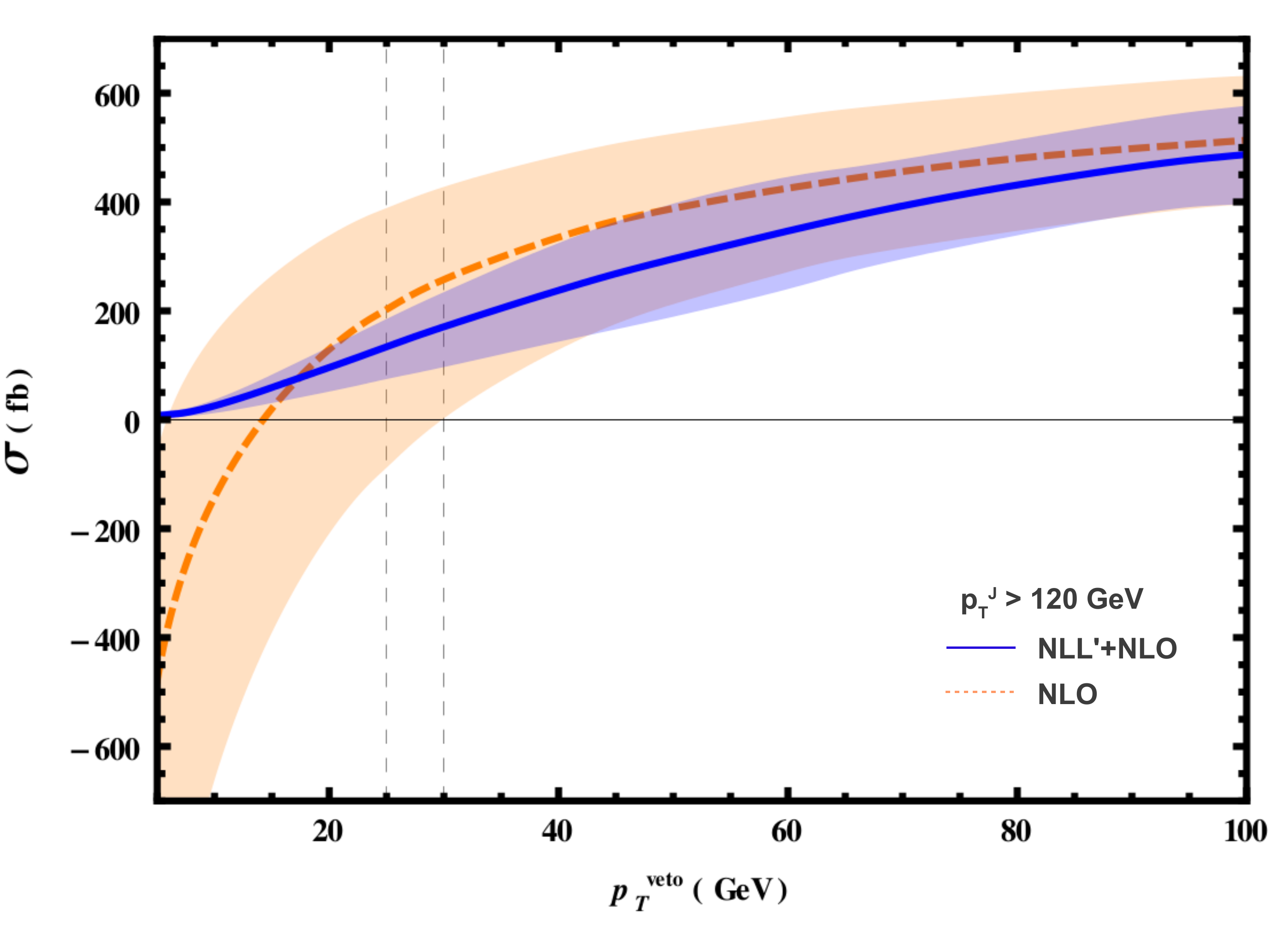}
\end{center}
\vspace{-0.5cm}
\caption{Presented here are the ${\rm NLO}$ v.s. ${\rm NLL}'+{\rm NLO}$ integrated cross sections with
 $p_T^J > 120\,{\rm GeV}$. The blue solid line
is for the RG-improved cross section and the
yellow dashed line is the NLO prediction . The narrow blue band is obtained using Eq.~(\ref{UNLL}) for the uncertainty after resummation, while the
wide yellow band comes from using Eq.~(\ref{UNLO}) for the fixed-order 
uncertainty. }\label{120ptj}
\end{figure}

We begin by showing in Fig.~\ref{120ptj} the NLO and $\text{NLL}^{\prime}+\text{NLO}$ results for $p_T^J  >120$ GeV as a function of $p_T^{veto}$, to make clear the improvement gained by adding the resummation in the $p_T^J \sim m_H$ region.  
After resummation, the scale dependence has been dramatically reduced
for small $p_T^{veto}$.  For large $p_T^{veto}$ the RG-improved cross section tends toward the fixed-order result, as desired.  The pathological behavior of the fixed-order cross section for low $p_T^{veto}$ is clear, as the central value becomes negative for $p_T^{veto} \approx 15$ GeV.  This pathology is removed in the $\text{NLL}^{\prime}+\text{NLO}$ result.

\begin{figure}[h!]
\begin{center}
  \includegraphics[width=3.5in,angle=0]{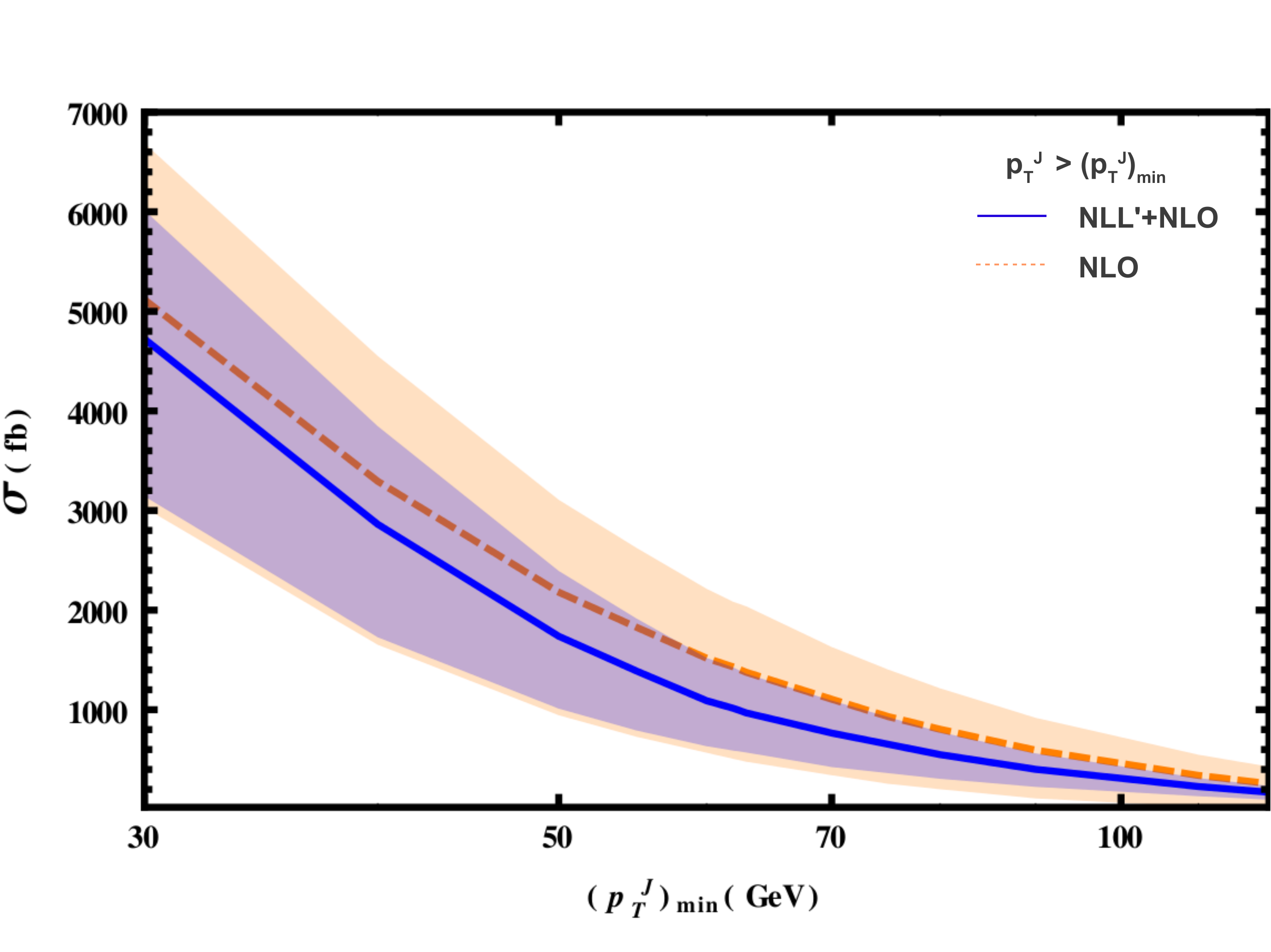}
\end{center}
\vspace{-0.5cm}
\caption{Shown are the ${\rm NLL}^{\prime}+ {\rm NLO}$ (blue band) and NLO (yellow band) cross sections for fixed $p_T^{veto} = 30\,{\rm GeV}$ as a function of the lower cut on $p_T^J$.}
\label{30ptveto}
\end{figure}

We continue by showing in Fig.~\ref{30ptveto} the cross section as a function of the lower cut on $p_T^J$ for $p_T^{veto}=30$ GeV.  Even for values of the lower $p_T^J$ cut near $p_T^{veto}$, a sizeable reduction of the uncertainty occurs when the 
${\rm NLL}^{\prime}+ {\rm NLO}$ result is used.  The reason for this is discussed in Sec.~\ref{sec:valid}; roughly half of the uncertainty comes from the high-$p_T^J$ region, which is exactly the parameter space improved by our effective-theory description.

Finally, we present in Table~\ref{tab:nums} numerical results for both the cross sections and the fraction of events in the one-jet bin, $f^{1j}$.  We define the event fraction as
\be
f^{1j}_x = \frac{\sigma_x}{\sigma_{inc}},
\ee
where $x$ denotes either the NLO or the $\text{NLL}^{\prime}+\text{NLO}$ cross section in the one-jet bin.  We note that our values for $f^{1j}_{NLO}$ are consistent with those obtained by the ATLAS collaboration~\cite{jianming2}, which provides a cross-check of our results.The total cross section $\sigma_{inc}$, as well as its estimated uncertainty, is taken from the LHC Higgs cross section working group~\cite{lhcxs}.  The uncertainties shown are calculated as discussed in Sec.~\ref{sec:uncert}.  Results are given for $m_H=124-126$ GeV, and for $p_T^{veto}=25$ and 30 GeV.  The reductions of the uncertainties are significant for both values of $p_T^{veto}$.  Symmetrizing the error for this discussion, the estimated uncertainty on the cross section improves from $\pm 40\%$ at NLO to $\pm 30\%$ at $\text{NLL}^{\prime}+\text{NLO}$, a reduction of one quarter of the initial value.  The one-jet fraction uncertainty decreases from $\pm 44\%$ to $\pm 34\%$.  For $p_T^{veto}=30$ GeV, the error on the cross section decreases from $\pm 36\%$ to $\pm 29\%$ when resummation is included, while the error on $f^{1j}$ decreases from $\pm 39\%$ to $\pm 33\%$.  Numerical results for other parameter choices are available from the authors upon request.  We note that these are extremely conservative error estimates, as discussed in Sec.~\ref{sec:uncert}.  We default to the ST prescription over a large region of the relevant parameter space, and turn off the resummation at a relatively high value of $p_T^J$.  Enough of the error comes from the high $p_T^J$ region that our RG-improvement is effective in taming the uncertainty.

\begin{table}
\begin{center}
\begin{tabular}{|c|c|c|c|c|c|}
\hline
$m_H$ (GeV) & $p_T^{veto}$ (GeV) & $\sigma_{\text{NLO}}$ (pb) & $\sigma_{\text{NLL}^{\prime}+\text{NLO}}$ (pb) &
	$f^{1j}_{\text{NLO}}$ & $f^{1j}_{\text{NLL}^{\prime}+\text{NLO}}$ \\
\hline \hline
124 & 25 & $5.92^{+35\%}_{-46\%}$ & $5.62^{+29\%}_{-30\%}$ & $0.299^{+38\%}_{-49\%}$ &  
	$0.283^{+33\%}_{-34\%}$ \\ \hline
125 & 25 & $5.85^{+34\%}_{-46\%}$ & $5.55^{+29\%}_{-30\%}$ & $0.300^{+37\%}_{-49\%}$ &  
	$0.284^{+33\%}_{-33\%}$ \\ \hline
126 & 25 & $5.75^{+35\%}_{-46\%}$ & $5.47^{+30\%}_{-30\%}$ & $0.300^{+38\%}_{-49\%}$ &  
	$0.284^{+34\%}_{-33\%}$ \\ \hline
\hline
124 & 30 & $5.25^{+31\%}_{-41\%}$ & $4.83^{+29\%}_{-29\%}$ & $0.265^{+35\%}_{-43\%}$ &  
	$0.244^{+33\%}_{-33\%}$ \\ \hline
125 & 30 & $5.19^{+32\%}_{-41\%}$ & $4.77^{+30\%}_{-29\%}$ & $0.266^{+35\%}_{-43\%}$ &  
	$0.244^{+33\%}_{-33\%}$ \\ \hline
126 & 30 & $5.12^{+32\%}_{-41\%}$ & $4.72^{+30\%}_{-29\%}$ & $0.266^{+35\%}_{-43\%}$ &  
	$0.246^{+33\%}_{-32\%}$ \\ \hline
\end{tabular}
\caption{Shown are the central values and uncertainties for the NLO cross section, the resummed cross section, and the event fractions in the one-jet bin using both the fixed-order and the resummed results.  Numbers are given for several Higgs masses and for $p_T^{veto}=25,30$ GeV.}
\label{tab:nums}
\end{center}
\end{table}

\section{Conclusions}\label{sec:conc}

We have studied in detail the resummation of a class of large Sudakov logarithms appearing in the perturbative expansion for Higgs production in the one-jet bin.  These logarithms occur when the jet transverse momentum is of order the Higgs mass.  For certain parameter choices they lead to pathological behavior of the fixed-order perturbative expansions, including negative cross sections and scale-uncertainty estimates that do no properly account for missing higher-order corrections.  Past attempts to handle this problem increased the theoretical error estimate.  While theoretically correct, this has the unfortunate effect of introducing a large systematic error into experimental analyses.  The results we present here 
tame the poor behavior of fixed-order QCD by controlling the jet-veto logarithms to all orders in perturbation theory, leading to a more reliable theoretical prediction and a reduced uncertainty estimate.  We have reviewed the necessary formalism to understand the resummation, and have extended the theoretical accuracy to the $\text{NLL}^{\prime}$ level by calculating the full one-loop soft function for this process.  Using our matched $\text{NLL}^{\prime}+\text{NLO}$ result for the cross section, we have performed a detailed numerical study of exclusive Higgs plus one-jet production at the LHC, including an estimation of the non-global logarithmic effect and a careful accounting of the theoretical uncertainties before and after resummation.  The estimated theoretical uncertainties in the one-jet bin are reduced by up to 25\% using the $\text{NLL}^{\prime}+\text{NLO}$ prediction.  This is an extremely conservative error estimate, as argued in the main text. 

We believe that it is now time to revisit the theoretical error treatment used by the LHC experiments in their studies of Higgs properties.  Current studies use fixed-order perturbation theory, and the error treatment suggested in Ref.~\cite{Stewart:2011cf}, to estimate the uncertainties induced by dividing the signal into exclusive jet multiplicites.  While this is the best choice to correctly handle the uncertainties induced by the jet veto if only fixed-order predictions are available, the resummation of jet-veto logarithms is now known for both the zero-jet and one-jet bins.  These predictions do not exhibit the pathological scale variation that led to the prescription of Ref.~\cite{Stewart:2011cf}.  We are confident that this is only the first of many upcoming improvements for predictions of Higgs-boson production in association with a fixed number of jets.  A new factorization theorem for the low-$p_T^J$ region for exclusive Higgs plus one-jet production, an improved treatment of $\text{ln} \,R$ effects, and new high-precision fixed-order calculations will all further reduce the theoretical uncertainties and help provide a sharpened image of the newly-discovered scalar at the LHC.  We encourage the experimental communities to incorporate this new knowledge into their analyses.

\section*{Acknowledgments}

We thank J. Qian and G. Salam for helpful discussions.  This work was supported by the U.S. Department of Energy, Division of High Energy Physics, under contract DE-AC02-06CH11357 and the grants DE-FG02-95ER40896 and DE-FG02-08ER4153.

\appendix
\section{Fixed-order jet and beam functions}
In this Appendix, we tabulate all ingredients needed for resummation at $\text{NLL}^{\prime}$ accuracy. We start
with the NLO calculation of the jet and the beam functions, whose operator definitions
can be found in Ref.~\cite{Liu:2012sz}.  The anti-$k_T$ jet function is calculated using the measurement function
\bea
\hat{\cal M}_J = \Theta(\Delta \eta_{ij}^2 + \Delta \phi_{ij}^2 < R^2 )
+{\cal O}(p_T^{veto})\,.
\eea
We explicitly check that the collinear radiation
leaking outside the jet is power-suppressed by $p_T^{veto}$ after correctly subtracting the soft zero-bin contributions.  Since numerically $R\ll 1$, we can simplify the measure using
\bea
\Delta\eta_{ij}^2 + \Delta \phi_{ij}^2 = 
2\cosh(\Delta\eta_{ij}  ) - 2\cos(\Delta \phi_{ij}) + {\cal O}(R^4).
\eea
In this limit, the NLO jet functions for gluons and quarks become
\bea
J^{(1)}_g &=& \,
\frac{\alpha_s(\mu)}{2\pi}\left[C_A\left(\frac{67}{9}-\frac{3\pi^2}{4} \right) \,
-\frac{23}{9}\frac{n_f}{2}\,
+\beta_0\log \frac{\mu}{p_T^JR}\,
+2C_A \log^2 \frac{\mu}{p_T^J R} \,
\right] +{\cal O}(R^2) \,, \nn \\
J^{(1)}_q &=& \,
\frac{\alpha_s(\mu)}{2\pi}C_F\,
\left[\frac{13}{2}-\frac{3\pi^2}{4}
+3\log \frac{\mu}{p_T^JR}\,
+2\log^2\frac{\mu}{p_T^J R} \,
\right] +{\cal O}(R^2) \,,
\eea
similar to what was obtained in Ref.~\cite{Ellis:2010rwa} using a slightly different jet
algorithm.
We note that the jet functions are normalized so that the leading-order results are unity for both quarks and 
gluons: $J_i^{(0)} = 1$.

The measurement operator for the beam function with a single emission is 
\bea
\hat{\cal M}_B = \Theta\left(k_{T,i} < p_T^{veto}\right)\, 
\Theta\left(|\eta_i| < \eta_{\rm cut} \right) \,
+ \Theta(|\eta_i| > \eta_{\rm cut} )\,.
\eea
Experimentally, $\eta_{\rm cut} \sim 4.5$.  For simplicity, we set
$\eta_{\rm cut} = \infty$ here.  We note that this difference does not affect the anomalous dimension of the beam function, it only changes the finite part. 
The calculation is performed using the 't Hooft-Veltmann scheme.
The NLO matching coefficient ${\cal I}$ for the various beam functions are found to be
\bea
{\cal I}^{(1)}_{gg}(z) &=& \,
\frac{\alpha_s(\mu)C_A}{2\pi}\left(\,
4\log\frac{\mu}{p_T^{veto}}\log\frac{\nu}{\bnp} \delta(1-z)
-2\tilde{p}_{gg}(z) \log \frac{\mu}{p_T^{veto}} 
 \right)\,, \nn \\
{\cal I}^{(1)}_{qq}(z) &=& \,
\frac{\alpha_s(\mu)C_F}{2\pi}\left(\,
4\log\frac{\mu}{p_T^{veto}}\log\frac{\nu}{\bnp}\delta(1-z)\,
-2\tilde{p}_{qq}(z) \log \frac{\mu}{p_T^{veto}}\,
+(1-z)
\right)\,, \nn \\
{\cal I}^{(1)}_{gq}(z) &=& \frac{\alpha_s(\mu)C_F}{2\pi}\left(\,
-2p_{gq}(z)\log\frac{\mu}{p_T^{veto}} + z 
\right)\,, \nn \\
{\cal I}^{(1)}_{qg}(z) &=& \frac{\alpha_s(\mu)T_F}{2\pi}\left(\,
-2p_{qg}(z)\log\frac{\mu}{p_T^{veto}} + 2z(1-z) 
\right) \,,
\eea
with 
\bea
&&\tilde{p}_{gg}(z) = \frac{2z}{(1-z)_+}+2z(1-z)+2\frac{1-z}{z}\,, \nn\\
&&\tilde{p}_{qq}(z) = \frac{1+z^2}{(1-z)_+}  \,, \nn \\
&&p_{gq}(z) = \frac{1+(1-z)^2}{z} \,, \nn \\
&&p_{qg}(z) = 1 -2z +2z^2\,.
\eea
We note that the beam functions are normalized so that the diagonal matching coefficients are delta-functions at tree-level, 
while the off-diagonal ones vanish: ${\cal I}_{ii}^{(0)} = \delta(1-z)$, ${\cal I}_{ij}^{(0)} = 0$ for $i \ne j$.

\section{Anomalous dimensions and RG evolution}

The beam, jet, soft and hard functions appearing in the factorization theorem of Eq.~(\ref{factgen}) all satisfy evolution 
equations of the form
\bea
\mu\frac{\mathrm{d}F}{\mathrm{d}\mu} = 
\Gamma^\mu_F(\mu) F(\mu)\,.
\eea
The soft and beam functions, which also contain rapidity divergences, have similar evolution equations in the rapidity regulator:
\bea
\nu\frac{\mathrm{d}F_{B,S}}{\mathrm{d}\nu} = 
\Gamma^\nu_{B,S}(\nu) F_{B,S}(\nu)\,.
\eea
These are easily extracted form the poles of the one-loop calculations for each of these objects.  The general solution to these RG equations can be formally written as
\bea
F(\mu,\nu) = U(\mu,\nu,\mu_0,\nu_0) F(\mu_0,\nu_0) \,.
\label{RGsols}
\eea
These RG-improved expressions for the beam, soft, jet and hard functions are then used in Eq.~(\ref{factgen}).  The initial conditions in Eq.~(\ref{RGsols}) can be determined from the fixed-order calculation presented in the previous section and in the main text for the soft function.

From these fixed order calculations, we can determine the anomalous dimensions
used for RG evolution. The anomalous dimension for the jet function is given by
\bea
\Gamma_{J_i} = 2\Gamma_{\rm cusp} 
T_i^2 \log \frac{\mu}{p_T^{J_i} R} +\gamma_{J_i}\,.
\eea
Here, $i$ labels the parton flavor, and can take on the values $i=q,g$.  For the beam function, the anomalous dimensions  are
\bea
\Gamma_B^\nu &=& 2\Gamma_{\rm cusp} T_i^2 \log \frac{\mu}{p_T^{veto}} \,, \nn\\
\Gamma_B^\mu &=& 2\Gamma_{\rm cusp} T_i^2 \log \frac{\nu}{\bnp} + \gamma_{B_i}\,,
\eea
where $T_i^2 = C_A$ for gluon and $T_i^2 = C_F $ for quark.  The anomalous dimensions can be obtained as expansions in the strong coupling constant.  For the resummation performed here, we need the following terms in each expansion:
\bea
\Gamma_{\rm cusp} &=& \frac{\alpha_s}{4\pi}\Gamma_0 
+ \left(\frac{\alpha_s}{4\pi}\right)^2\Gamma_1 + \dots , \nn \\
\gamma_{B_i} &=& \frac{\alpha_s}{4\pi}\gamma_0^i, \;\; \gamma_{J_i} = \frac{\alpha_s}{4\pi}\gamma_0^i .
\eea
We note that the non-cusp anomalous dimensions of the beam and jet functions are the same. We have the following expressions for the necessary anomalous dimensions, as well as the relevant coefficients of the QCD beta functions needed both here and later:
\bea
&&\beta_0 = \frac{11}{3} C_A - \frac{4}{3} T_F n_f  \,,\nn \\
&&\beta_1 = \frac{34}{3} C_A^2 - \frac{20}{3} C_A T_F n_f - 4 C_F T_F n_f \, , \nn \\
&&\Gamma_0 = 4 \,,\nn \\
&&\Gamma_1 = 4 \left[ C_A \left( \frac{67}{9} - \frac{\pi^2}{3} \right) -
\frac{20}{9} T_F n_f \right]\,, \nn \\
&&\gamma_0^g = 2 \beta_0, \;\;\; \gamma_0^q = 6 C_F.
\eea
The RG evolution of the soft function can be easily obtained by direct differentiation of Eq.~(\ref{finalsoft}).  As the soft function is sensitive to both the jet and beam directions, its anomalous dimensions take on a more complicated form than those of the other quantities.  We find
\bea
\Gamma^\mu_S &=& 2 \Gamma_{\rm cusp} \left\{ (T_a^2+T_b^2) \, \text{ln}\frac{\mu}{\nu} +y_J (T_a \cdot T_J - T_b \cdot T_J) \, \text{ln}\frac{\mu}{p_T^{veto}}\right\}, \nn \\
\Gamma^\nu_S &=& -2 \Gamma_{\rm cusp} \left(T_a^2+T_b^2\right) \, \text{ln}\frac{\mu}{p_T^{veto}},
\eea
where $a,b$ denote the beam directions and $J$ the final-state jet direction.  We note that $\Gamma^\nu_S+\Gamma_{B_a}^\nu+\Gamma_{B_b}^\nu =0$, as required by RG-invariance of the cross section under $\nu$-variations.  The anomalous dimension of the hard function was studied in detail in Ref.~\cite{Kelley:2010fn}, and we do not reproduce it here.  It can be derived from the above expression using RG-invariance of the cross section: $\Gamma^\mu_S+\Gamma_{B_a}^\mu+\Gamma_{B_b}^\mu+\Gamma_J^{\mu}+\Gamma_H^{\mu} =0$.

\section{Solutions of the RG equations}

We reproduce here the solutions to the RG equations presented in the previous section, which are required in Eq.~(\ref{RGsols}).  The evolutions of the jet and the beam functions are given by
\bea
U_{J_i}(\mu_J,\mu) &=& \exp\left[-2T_i^2 S(\mu_J,\mu)-A_{J_i}(\mu_J,\mu) \right]
\left( \frac{\mu_J}{p_T^{J}R}\right)^{-2T_J^2 A_\Gamma(\mu_J,\mu)} \,, \nn \\
U_{B,a}(\mu_B,\nu_B,\mu,\nu) &=& 
\exp\left[-T_a^2 A_\Gamma(p_T^{veto},\mu)\log\frac{\nu^2}{\nu_B^2}\right]
\exp\left[-T_a^2 A_\Gamma(\mu_B,\mu)\log\frac{\nu^2_B}{\w_a^2}-A_{B_a}(\mu_B,\mu) \right]\,. \nn \\
\eea
The solution to the RG equation for the hard function is
is
\bea
U_H (\mu_H,\mu)&=&\exp\left[2\sum_i T_i^2 S(\mu_H,\mu)-2A_H(\mu_H,\mu)\,
+2A_\Gamma(\mu_H,\mu) \sum_{i\neq j}\frac{T_i\cdot T_j}{2}\log\Delta R^2_{ij}\right]\nn\\
&& \times
 \prod_i \left(\frac{\mu_H}{\w_i} \right)^{2T_i^2A_\Gamma(\mu_H,\mu)}\,, 
\eea
where we have set $\Delta R^2_{Ja} = e^{-\eta_J}$, $\Delta R^2_{Jb} = e^{\eta_J}$ and $\Delta R^2_{ab} = 1$. We also set $\w_i = p_T^{J}$ if $i = J$, otherwise we have $\w_a = x_a\sqrt{s}$.  The soft-function evolution factor is
\bea
U_S(\mu_S,\nu_S,\mu,\nu)& = &\exp\left[-2\sum_{i\in B}T_i^2 S(\mu_s,\mu) -A_s(\mu_s,\mu)\, 
- 2A_\Gamma(\mu_s,\mu)\sum_{i \ne j}\frac{T_i\cdot T_j}{2}\log \Delta R^2_{ij}\right]\nn\\
&& \times \left(\frac{1}{R}\right)^{2T_J^2A_{\Gamma}(\mu_s,\mu)}
\left(\frac{\nu_s}{\mu_s} \right)^{\sum_{i\in B}2T_i^2A_{\Gamma}(\mu_s,\mu)}
\left(\frac{\nu}{\nu_s} \right)^{\sum_{i\in B}2T_i^2A_{\Gamma}(p_T^{veto},\mu)}\,.
\eea
For the NLL resummation, we need the following factors:
\bea
A_{\Gamma}(\mu_i,\mu_f)&=& \frac{\Gamma_0}{2\beta_0}\left\{
\log r + \,
\frac{\alpha_s(\mu_i)}{4\pi}\left(\frac{\Gamma_1}{\Gamma_0}-\frac{\beta_1}{\beta_0} \right)\,
(r-1) \,
\right\}\,,
\eea
and 
\bea
S(\mu_i,\mu_f) &=& \frac{\Gamma_0}{4\beta_0^2}\left\{ 
\frac{4\pi}{\alpha_s(\mu_i)}\left(1-\frac{1}{r}-\log r\right)\,
+\left(\frac{\Gamma_1}{\Gamma_0}-\frac{\beta_1}{\beta_0} \right)(1-r+\log r)\,
+\frac{\beta_1}{2\beta_0}\log^2 r 
\right\}\,, \nn \\
\eea
where $r=\alpha_s(\mu_f)/\alpha_s(\mu_i)$. 
$A_{J/B}$, $A_H$ and $A_S$ are needed at leading order, and can be obtained
by substituting the $\Gamma_0$ in $A_\Gamma$ with the corresponding
$\gamma_0^i$ and expanding in $\alpha_s$.

\end{document}